\documentclass[dvipsnames,conference]{IEEEtran}
\usepackage{courier}
\usepackage[ruled,vlined]{algorithm2e}
\usepackage{graphicx}
\usepackage{paralist}
\usepackage[shortlabels]{enumitem}
\usepackage{tabularx}
\usepackage{balance}
\usepackage{multirow}
\usepackage{amsfonts,amssymb}
\usepackage{amsthm}
\usepackage{multicol}
\usepackage{pbox}
\usepackage{mathtools}
\usepackage[TABBOTCAP]{subfigure}
\usepackage{algorithmic}
\usepackage{paralist}
\usepackage{tabularx}
\usepackage{url}
\usepackage{balance}
\usepackage{multirow}
\usepackage{multicol}
\usepackage{setspace}
\usepackage{enumitem}
\usepackage{verbatim}
\usepackage{float}
\usepackage[normalem]{ulem}
\usepackage{xspace}
\usepackage{ifthen}
\usepackage{bold-extra}
\usepackage{hyperref}
\usepackage{pifont}
\usepackage{tikz}
\usepackage{xcolor}
\usepackage{enumitem}
\usepackage{array}
\usepackage{multirow}

\usepackage{booktabs}

\usepackage{graphicx}

\input{macro}

\newcommand\MyBox[2]{
  \fbox{\lower0.75cm
    \vbox to 1.7cm{\vfil
      \hbox to 1.7cm{\hfil\parbox{1.4cm}{#1\\#2}\hfil}
      \vfil}%
  }%
}

\newtheorem{definition}{Definition}

\definecolor{MidnightBlue}{HTML}{01693F}

\newcommand{\FRAME}{{\sc Frame}\xspace}

\newcommand{\arun}[1]{{#1}}

\begin{document}

\title{Fusing UI Structure \& Semantics for Feature-Oriented App Screen Retrieval \& Clustering}

\author{
\IEEEauthorblockN{
Arun Krishna Vajjala\IEEEauthorrefmark{1},
Yanfu Yan\IEEEauthorrefmark{2},
Ajay Krishna Vajjala\IEEEauthorrefmark{1},\\
Shrunal Pothagoni\IEEEauthorrefmark{1},
Denys Poshyvanyk\IEEEauthorrefmark{3},
Kevin Moran\IEEEauthorrefmark{4}
}

\IEEEauthorblockA{\IEEEauthorrefmark{1}
George Mason University, Fairfax, VA, USA, \{akrishn, akrish, spothago\}@gmu.edu
}

\IEEEauthorblockA{\IEEEauthorrefmark{2}
Zhejiang University, Hangzhou, China, yanfu@zju.edu.cn
}

\IEEEauthorblockA{\IEEEauthorrefmark{3}
William \& Mary, Williamsburg, VA, USA, dposhyvanyk@wm.edu
}

\IEEEauthorblockA{\IEEEauthorrefmark{4}
University of Central Florida, Orlando, FL, USA, kpmoran@ucf.edu
}
}

\maketitle

\begingroup
\renewcommand\thefootnote{}
\footnotetext{\IEEEauthorrefmark{2}Corresponding author.}
\endgroup

\begin{abstract}

User Interface (UI) programming is challenging due to the complex abstraction gap between code and graphical software representations. To bridge this gap, UI programming tools often rely on screen retrieval and clustering, which require accurate similarity measures based on overlapping features. However, computing feature-oriented similarity is difficult because screens with similar functionality often exhibit design variations.

To address this, we propose \FRAME (Rein\textbf{\underline{F}}orced Use\textbf{\underline{R}} Interf\textbf{\underline{A}}ce Screen E\textbf{\underline{M}}bedding with Graphical Structural Compr\textbf{\underline{E}}hension), a multi-modal, neuro-symbolic embedding technique. \FRAME constructs symbolic, graph-based representations of UI components to encode salient relationships and capture feature patterns across different screens. It leverages large vision-language models for visual and lexical encoding, alongside a novel UI-specific computational geometry algorithm that enables weighted embedding propagation. Across three benchmarks, \FRAME outperforms strong baselines by up to 13\% MRR in search and 7.6 percentage points in clustering accuracy. A comprehensive ablation study further confirms the benefit of each component, demonstrating \FRAME's potential for enhancing automated UI design and testing tools.

\end{abstract}


\section{Introduction}
\label{sec:introduction}

The engineering process behind building, testing, and maintaining high-quality user interfaces (UIs) has been long documented to be uniquely challenging~\cite{Myers:CHD94}.
UI programming encompasses several obstacles related to difficulties reasoning about event-based programming, low-testability, limited language support, and programming tool complexity~\cite{Myers:CHD94,Myers:CHI92,Samudio:VLHCC22}. The key underlying challenge that leads to these obstacles is the need for programmers to navigate a complex abstraction gap, reasoning across \textit{three} modalities of information: (i) natural language (documentation, requirements, etc.), (ii) code, and (iii) the pixel-based, graphical representation of the UI itself. This reasoning process can be particularly challenging as UI programming tends to follow paradigms which have been illustrated to be difficult to learn and understand, even for experienced programmers~\cite{Rosson:CHI:87,Tucker:CSH04,Moran:ICSE'18,Samudio:VLHCC22}.

To help overcome these challenges, the research community has invested in creating tools for automating UI design and programming tasks. Underlying many of these tools are the operations of screen \textit{retrieval} and \textit{clustering}, which, given a query screen retrieve semantically similar screens from a larger corpus or cluster semantically similar screens together, respectively. For example, recent programming tools related to UI testing rely on screen similarity to perform test transfer~\cite{Avgust}, enable context specific UI testing scenarios~\cite{khan2024aurora}, uncover non-crashing bugs~\cite{Su:FSE22}, and make UI testing more efficient~\cite{Feng:ICSE'23}.  UI programming tools related to bug report management use screen similarity to identify duplicate UI-based bug reports~\cite{Yan:ICSE'24} and perform fault localization~\cite{saha2024screenbug}. Finally, screen similarity is also important in UI design and implementation to assist in UI prototyping~\cite{Moran:TSE'18,Behrang_18} and evolution~\cite{Salma:MSR'24}. Given the increasing impact that models for screen retrieval and clustering are having on modern UI programming tools, their accuracy, generalizability, and robustness are critical.

In both retrieval and clustering tasks, the core mechanic is rooted in computing \textit{screen similarity} according to the semantics, or meaning, of the screens. In the context of UI programming tools screen semantics are generally defined by the features afforded or supported by the UI. As such in this paper, we define the phrase \textit{semantic screen similarity} to refer to the extent to which a pair of UI screens share similar supported features. Capturing semantic screen similarity can be exceedingly challenging given the wide variations in UI designs that support similar features~\cite{Rico}.

\begin{figure}[t]
    \centering
    \includegraphics[width=0.84\columnwidth]{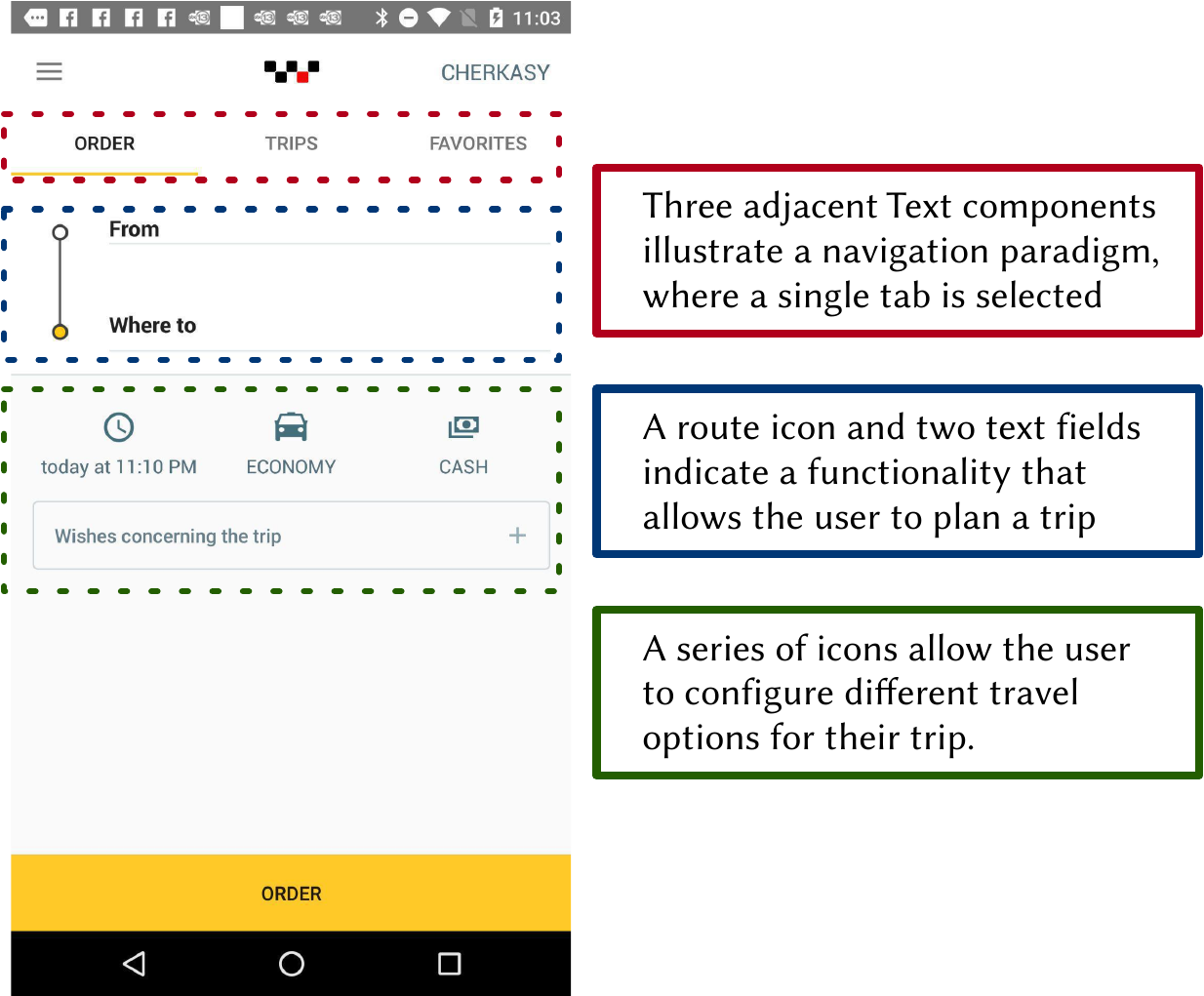}
    \caption{Example UI Component Relationships \& Corresponding Features}
    \label{fig:screen-ex}
\end{figure}

Existing methods of measuring screen similarity may capture some visual, lexical, or even structural context, but they generally fall short in representing \textit{the relationships between UI components}—relationships that are essential to understanding screen features and functionality. For instance, grouped components like navigation menus provide a screen’s functional layout and meaning, which many current models do not adequately capture. Nuanced relationships between the elements on a given screen can provide important context on supported features in screen retrieval and clustering tasks. Figure~\ref{fig:screen-ex} illustrates the importance of these relationships, wherein the components grouped in red signify navigation, the components grouped in blue signify a trip planning widget, and the components grouped in green signify trip options.

In other words, there exist key limitations in current UI modeling techniques that hinder their effectiveness:
\arun{(i) \textit{Lack of Representation of Screen Features: } While general-purpose VLMs treat UI screens as monolithic pixel-grids~\cite{Li21,li2021vut,bai2021uibert}, SE tasks require reasoning about functional modules, groups of components that work together to provide a feature.} (ii) \textit{Expensive, proprietary, and domain-specific training procedures:} most current UI models require expensive training on UI-specific datasets that are difficult to source and label, and quickly become outdated~\cite{li2021vut,bai2021uibert}; and (iii) \textit{Lack of Shared UI Component Context:} Current models are limited in their ability to model joint context across related UI components.

To address the limitations of existing techniques for computational UI understanding of screen similarity,
we propose \FRAME: A Rein\textbf{\underline{F}}orced Use\textbf{\underline{R}} Interf\textbf{\underline{A}}ce Screen E\textbf{\underline{M}}bedding With Graphical Structural Compr\textbf{\underline{E}}hension. 
Our \textit{key insight} is that semantically coherent representations of UIs should follow a neuro-symbolic approach, leveraging the representational power of large vision-language models as a general-purpose information backbone, while explicitly modeling UI features by capturing widget relationships using a symbolic, graph-based structure. \FRAME embeddings function as an \textit{augmentation} of the embeddings from large vision-language models (VLMs) (\ie CLiP, BERT) and model UI relationships using computational geometry and a knowledge propagation algorithm, while automatically weighting the importance of UI component relationships -- and hence features. Because \FRAME is an \textit{augmentation} or extension of existing VLM embeddings, it can benefit from the continued improvement in the representation power of these models, \textit{and does not require any further training of VLMs}. Instead it functions by imposing strong symbolic priors using human knowledge of how features are constructed. In summary, \FRAME overcomes limitations in prior work by transforming ``flat'' pixel-level data into a structured graph that reflects the developer's conceptual model of UI functionality by augmenting VLM embeddings without any additional training expense.

To illustrate the effectiveness of \FRAME's embedding technique, we apply it to the tasks of screen retrieval and clustering, using three benchmark datasets derived from past software engineering tools supporting UI testing~\cite{khan2024aurora,Avgust}, and design~\cite{Leiva20_enrico}.
When compared to both strong generalized and UI-specific model baselines, \FRAME achieves up to 13\% higher MRR for screen retrieval and 7.6 percentage points higher accuracy in clustering, illustrating how our technique can improve the UI programming tools that our datasets originally supported.

\noindent The main contributions of this paper are as follows: \\
\begin{enumerate}[topsep=0pt,itemsep=0ex,partopsep=0ex]
    \item{A UI-aware, spatial knowledge propagation algorithm for encoding the relationship of UI components into a multimodal embedding;}
    \item{A weighting strategy for UI-relationships based on the Vietoris-Rips complex from computational geometry;}
    \item{\FRAME~-- an implementation of the above strategies in conjunction with VLMs for calculating the feature-based similarity of UI screens;}
    \item The results of a comprehensive empirical evaluation of \FRAME in the tasks of both screen retrieval and screen clustering across three SE-specific datasets;
	\item An online appendix~\cite{REPO,REPO-GH} that contains \FRAME's code, our evaluation data, and our experimental infrastructure to foster replicability.%
\end{enumerate}

\section{Constructing and Measuring Feature-Based Similarity with \FRAME Embeddings}
\label{sec:approach}
\vspace{-0.2em}

\begin{figure*}[t]
    \centering
    \vspace{-1em}
    \includegraphics[width=0.9\textwidth]{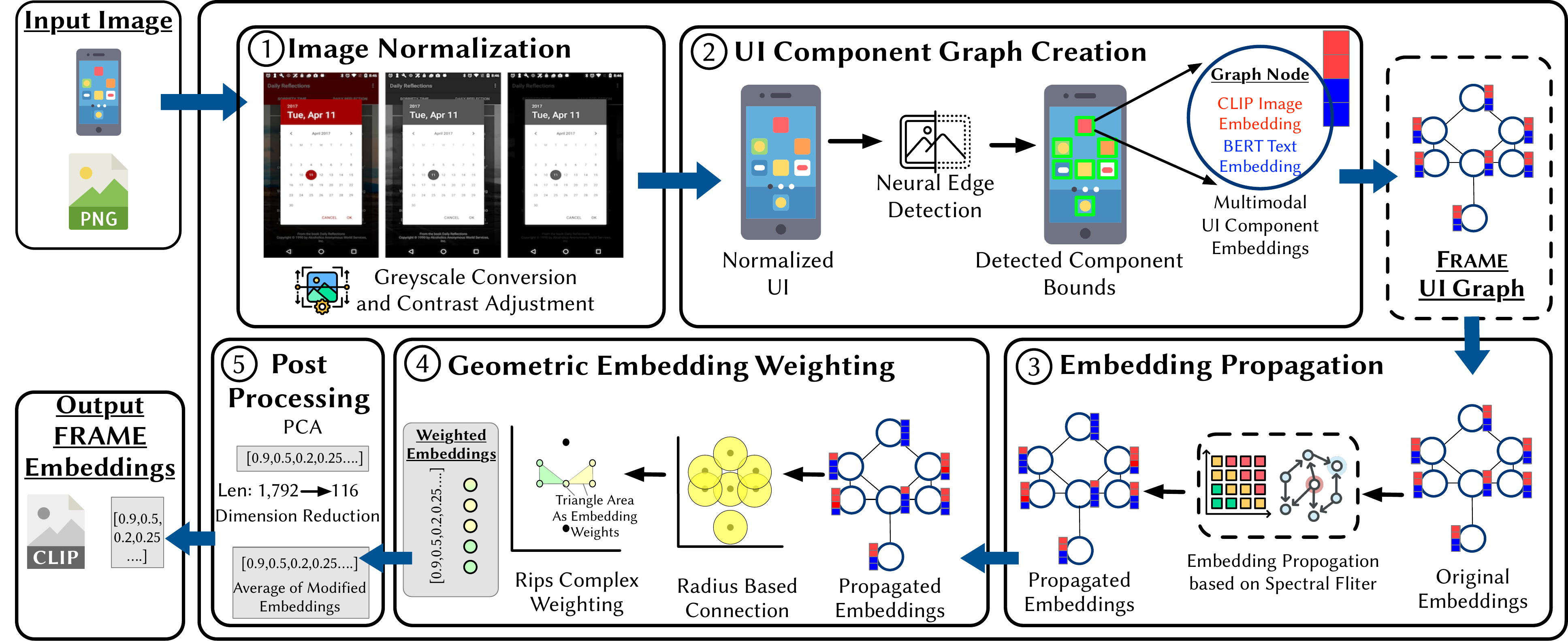}
    \vspace{-1em}
    \caption{Overview of the \FRAME Screen Embedding Workflow}
    \label{fig:overview}
    \vspace{-0.5em}
\end{figure*}

\FRAME encodes semantic feature-based properties into UI embeddings which can benefit UI programming tools that rely upon screen similarity calculations. To achieve this, \FRAME requires only the screenshot of a mobile app screen to generate the embedding, making it more easily extensible to a variety of contexts and UI programming tools. In this paper, we implement and evaluate \FRAME for mobile app screens, given their popularity. However, the general workflow is extensible to a diverse set of UI screens. 

The key idea underlying \FRAME is that semantic screen understanding requires a neuro-symbolic approach. That is, while generalized vectors of images and text derived from VLMs may be able to capture certain properties of UI screens, in order to measure a faithful \textit{feature-based} screen similarity, we need to encode aspects of UI structure, through widget relationships, given its role in defining features. \FRAME accomplishes this by encoding UI widgets as a graph, and using an embedding propagation technique in conjunction with a spatial weighting technique adapted from computational geometry to capture widget relationships. It is important to note that \FRAME is a significant \textit{augmentation} for VLM embeddings, requires no additional training whatsoever, and hence can improve over time as these base models improve.

\FRAME operates in five main stages to create a semantically accurate representation of the screen, illustrated in Figure~\ref{fig:overview}. \circled{1} First, in the \textit{Image Normalization Phase}, \FRAME performs preprocessing on a given input screenshot image, adjusting its colors and contrast to emphasize the structural boundaries of UI components. \circled{2} Next, in the \textit{UI Component Graph Creation} phase, \FRAME uses a neural object detection technique to localize the boundaries of UI components from the screenshot to create UI component nodes, and then both visual and lexical embeddings for each node are computed, and spatial connections between nodes are constructed. \circled{3} In the \textit{Embedding Propagation} phase \FRAME shares information across connected UI graph components using an approximation of the spectral filter of the constructed UI graph. \circled{4} In the \textit{Geometric Embedding Weighting} phase, groups of components embeddings are weighted by calculating the Vietoris-Rips complex~\cite{gudhi:RipsComplex}, a technique adapted from computational geometry, using spatial UI centroids. This allows us to capture spatial relationships between components that signify screen features. \circled{5} Finally, in the \textit{Post-Processing} phase, \FRAME uses Principal Component Analysis (PCA) to reduce the dimensionality of the full screen embeddings while preserving the most salient portions of the embedding. In the remainder of this section we describe each phase of \FRAME.

\subsection{Image Normalization}
In order for \FRAME's embedding to learn UI relationships without being biased by nuanced stylistic details it utilizes two image preprocessing techniques on a given input screenshot. Screens that offer near functional equivalence can differ widely across stylistic properties, including color palettes, customized icon designs, and tailored fonts. To mitigate the effects of these variations and learn transferable semantic patterns, \FRAME converts screens to greyscale and enhances screenshot contrast. This has the additional benefit of making image-based UI component localization in Phase-\circled{2} of \FRAME's workflow more effective.

\begin{figure}[t]
    \centering
    \includegraphics[width=\columnwidth]{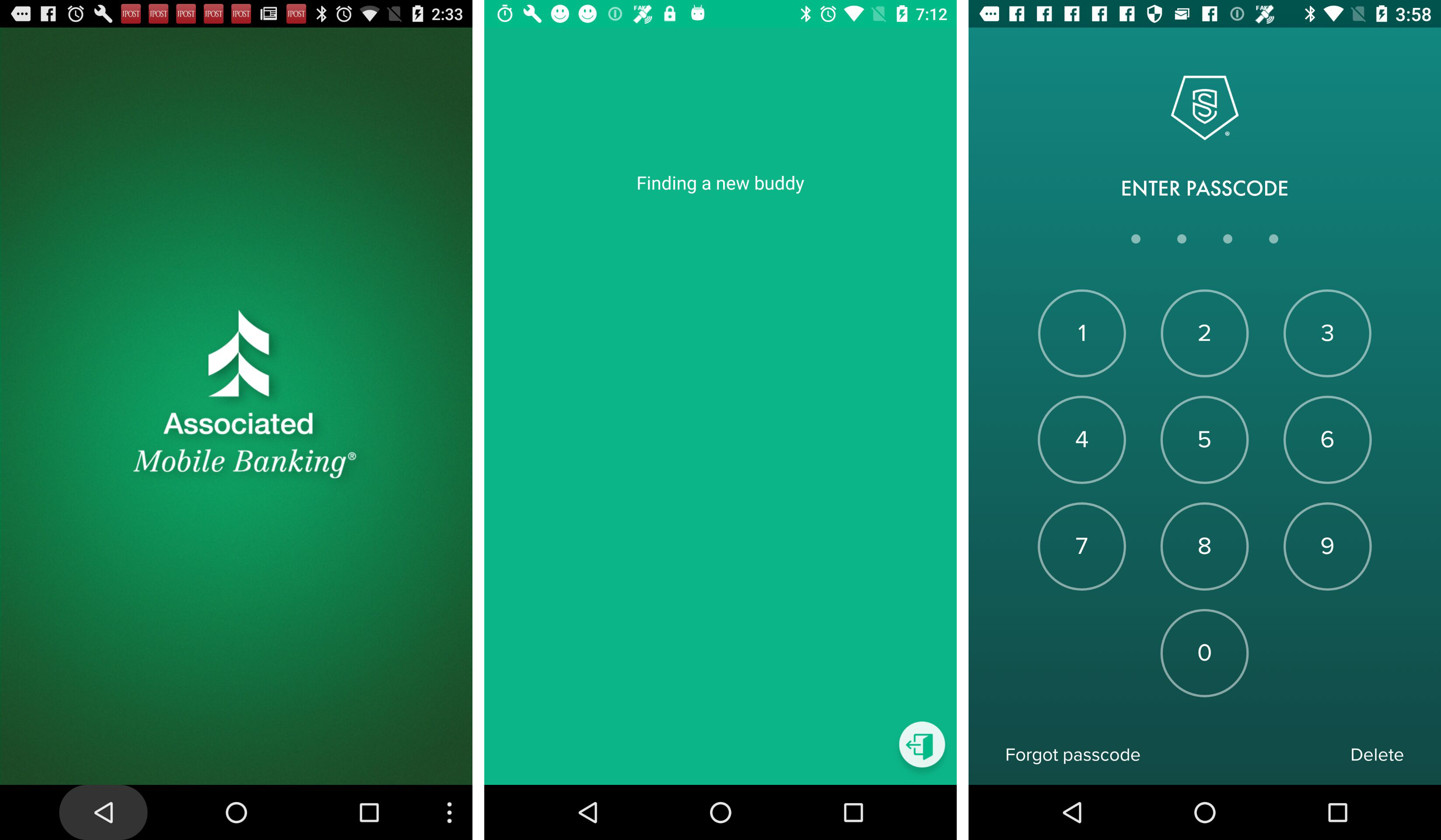}
    \caption{Three Images with Similar CLIP Embeddings}
    \label{fig:3clip}
\end{figure}

\subsubsection{Greyscale Conversion}
\FRAME's goal is to provide similarity rooted in supported screen features, rather than stylistic UI patterns. We illustrate how open-domain vision-language models tend to perceive UI screens in Fig.~\ref{fig:3clip}, which depicts three images from one of our evaluation datasets that were the most similar when using a CLIP embedding~\cite{clip}. It is clear that the three screenshots have very similar color palettes, yet they do not share the same structural components and represent three varying sets of functionality (or lack thereof). The leftmost screen is a Splash screen, the middle is a Search screen, and the rightmost screen is a password screen. Visual intuition suggests that these three screens provide different functions, however, because color plays a dominant role in VLM embeddings, such as those from CLIP, they are rated as similar. To reduce the seemingly large impact that color plays in VLMs we convert screens to greyscale, which preserve gradients that capture emphasis. \FRAME uses the Pillow library's~\cite{pillow} greyscale conversion to accomplish this.

\subsubsection{Contrast Adjustment} 

In addition to converting the input screenshot to greyscale, \FRAME alters the images by increasing the contrast by $2\times$ in the image. Contrast allows \FRAME to darken the dark pixels in the screen and brighten the brighter pixels in the screen. The resulting image has two main characteristics: (i) enhanced edge visibility; (ii) reduced noise influence -- both of which aid in visual UI component detection.
We use the Pillow library's~\cite{pillow} ``ImageEnhance'' functionality to double the contrast. An example is shown in Figure~\ref{fig:overview}-\circled{1}.%

\subsection{UI Component Graph Creation}

To encode the visual structure of a UI screen into a symbolic representation, \FRAME builds a graph by plotting nodes according to the absolute spatial coordinates of detected components on the screen. First, we adapt computer vision and deep learning techniques to identify UI components. Second, visual and lexical embeddings are created for each individual UI component, and edges are established via spatial relationships between components.

\subsubsection{UI Component Detection}

\FRAME adapts prior work that uses a combination of classical and neural computer vision techniques to detect UI components on a given screen~\cite{UIED}. We adapt these techniques to extract the bounding boxes of each UI component on a target screen which \FRAME uses to crop out an image of each component from the larger screenshot.
\FRAME only considers components larger than $48\times48$px since that is the standard for minimum accessible interactive icons on an Android screen as per Google's Accessibility Guidelines~\cite{GoogleAccess} -- this effectively allows us to filter out smaller elements that might dilute the learned UI component relationships. The centroid of each bounding box is used to signify each component's absolute location on the screen.%

\subsubsection{UI Graph Creation}

Given that both the visual design (\eg a search icon) and lexical information (\eg text displayed by a component) contribute to UI meaning, \FRAME encodes both types of information for each extracted UI component. To do this, \FRAME uses the cropped icon provided from the UI Component Detection procedure and creates two embeddings from it, an image embedding using a large vision-language model, and a text embedding using a large language model (LLM). In our implementation of \FRAME we elected to use CLIP~\cite{clip} embeddings to encode visual component-level information and BERT~\cite{BERT} to encode lexical information extracted via Optical Character Recognition (OCR) from individual component nodes in the UI graph. Given the number of embeddings that must be calculated for a given screen, a balance of inference efficiency, and the representative power of the model is our key concern. As such, we chose to use the more efficient BERT and CLIP models for text and image embeddings respectively since the token sequences present on UI components are typically short, CLIP has been shown adept at component modeling in the past~\cite{saha2024screenbug}, and the increased representational power of larger GPT-like models does not outweigh the efficiency of ``smaller'' models when embedding text/visuals from large number of UI components.

\underline{\textit{Image Embedding:}}
As shown in Fig.~\ref{fig:overview}, each node has a visual representation of the UI component that is cropped out of the screen. We pass each image to the Huggingface's open-source implementation of the CLIP model~\cite{CLIP-hug}, saving the resulting embedding vector.
 
\underline{\textit{Text Embedding:}} We use PyTesseract~\cite{Pytesseract} to extract text from each cropped UI component. We then pre-process the extracted text to remove punctuation and non-text/numeric characters. \FRAME uses this ``cleaned'' text to create the BERT embedding for each node.

With the inclusion of the image and text embeddings within each UI component, the nodes in the graph contain a multimodal data representation that is localized to each component. This gives each node a unique set of features that sets it apart from other nodes within the graph and its neighbors.

\underline{\textit{Graph Edges:}} Once nodes for the graph have been created, \FRAME constructs a graph by connecting nodes that are a maximum of 300 pixels in distance from each other according to the Manhattan distance measure. This number of 300 pixels was arrived at through experimentation on a held-out development set of screens from the RICO~\cite{Rico} dataset, where a systematic grid search of numbers were trialed and the resulting graphs visually inspected by two authors. Through this preliminary study, the two authors concluded that, for the largest majority of screen types, 300px was the most suitable for constructing semantically meaningful graph edges for mobile apps screens with a resolution of $1080\times1920$. For other screen sizes, using a distance roughly equivalent to 1/3 the screen width is likely suitable given our experimental observations. This approach allows for the formation of  graph neighborhoods that represent meaningful screen features.%

\subsection{UI Graph Embedding Propagation}
Unlike ordinary images, UI screens exhibit distinct component structures which encode information about the types of features afforded by the interface, with more pronounced relationships existing between spatially adjacent components. Therefore, we augment the image/text embedding of each component by its neighbor component embeddings based on the constructed UI graph edges from the prior phase.
Specifically, our embedding propagation strategy is derived from the first-order approximation of localized spectral filters on graphs~\cite{kipf2016semi, defferrard2016convolutional}, which can be represented as follows: 
\begin{equation} 
    S' = (I_N + w D ^ {-\frac{1}{2}} A  D ^ {-\frac{1}{2}}) S. 
\end{equation} 

\noindent $S \in \mathbb{R}^{N \times M}$ represents the matrix of all the image/text embeddings of nodes from $G$, and $S'$ represents the updated matrix by incorporating the information from the neighbor nodes. $M$ denotes the dimension of each image/text embedding (\eg $768$) and $N$ denotes the number of nodes in the graph. $A$ is the adjacency matrix of $G$ without self-connections and $D$ is the degree matrix of $A$ so that the adjacency matrix is normalized by $D$ with respect to both the row and the column. $w$ is a constant to balance the information from the original node/component with structural information from the neighbor nodes/components. By leveraging the embedding propagation strategy, we create shared embeddings wherein features from adjacent nodes are mixed, which serves as means to capture the relationships between UI components that signify features.  A visual representation of this process is shown in Fig.~\ref{fig:overview}-\circled{3}.

\vspace{-0.5em}
\subsection{Geometric Embedding Weighting}
\vspace{-0.5em}

For a given UI screen, not all spatial relationships among components are created equal. For example, encoding the three menu bar headers in the travel app presented at the outset of this paper (Figure~\ref{fig:screen-ex}) would be more important than encoding a relationship between one text title and the time picker widget. However, weighting the graph using spatial information is a complex task, which has been studied in topology in the field of mathematics. Simple weighting schemes using direct neighbors may not result in the creation of optimal topological weighting schemes that capture UI components indicating screen features. As such, to properly capture information regarding the topological relationship of various UI components and weight component relationships appropriately, \FRAME develops a new technique that applies concepts from computational geometry to UI data. The end result of this process is that components that would more typically signify UI features are grouped together, and those that do not remain separate.

\begin{figure}[t]
    \centering
    \vspace{-1em}
    \includegraphics[width=\columnwidth]{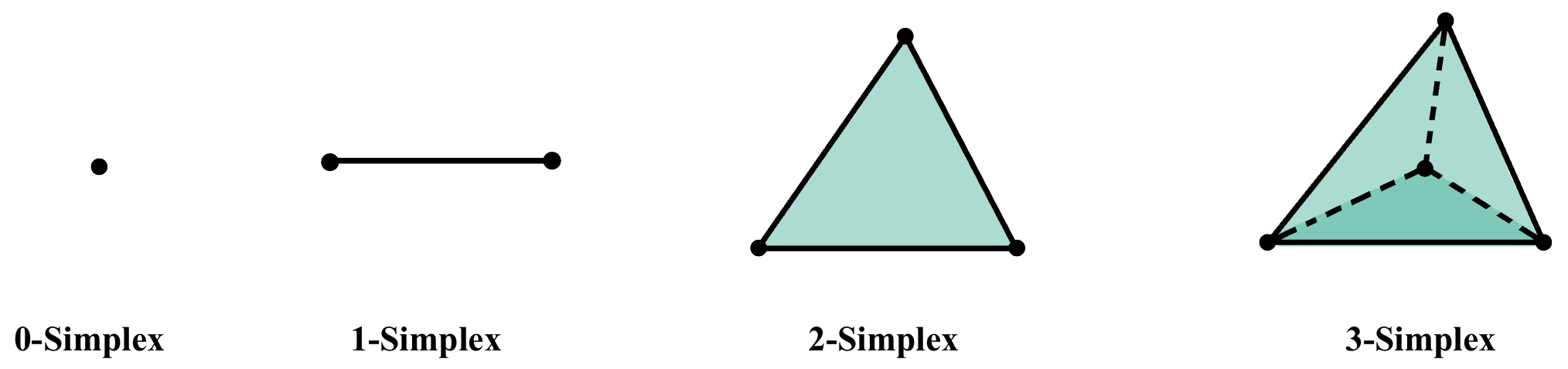}
    \caption{Visualization of Low-Dimensional Simplices}
    \label{fig:simplex}
\end{figure}

To do this, \FRAME represents the propagated UI embeddings as points in Euclidean space by triangulating the points to create a structure known as a simplicial complex (Fig.~\ref{fig:simplex}). 

These plotted embeddings are used to create an object known as the Vietoris-Rips complex (often called the Rips complex) using GUDHI~\cite{gudhi:RipsComplex}. Given a parameter $\epsilon>0$, the Rips complex of embeddings $\mathcal{X}$ is made into a simplicial complex that captures the spatial and topological relationships between the points in space using the following definition: \begin{definition} Given a set of embeddings $\mathcal{X} = \{x_1, x_2, \hdots, x_k\} \subset \R^{n}$ and an $\epsilon>0$, a $k$-simplex $\sigma = [x_{i_1}, x_{i_2}, \hdots x_{i_k}]$ is in the Vietoris-Rips Complex $\textit{Rips\,}_{\epsilon}(\mathcal{X})$ if and only if:
\[ \mathbb{B}_{\epsilon}(x_{i_j}) \cap \mathbb{B}_{\epsilon}(x_{i_{j'}}) \neq \emptyset \]
where $\mathbb{B}_{\epsilon}(x_{i_j})$ is an open ball of radius $\epsilon$ centered at $x_{i_j}$.
\end{definition}

\begin{figure}[t]
    \centering
    \vspace{-1em}
    \includegraphics[width=\columnwidth]{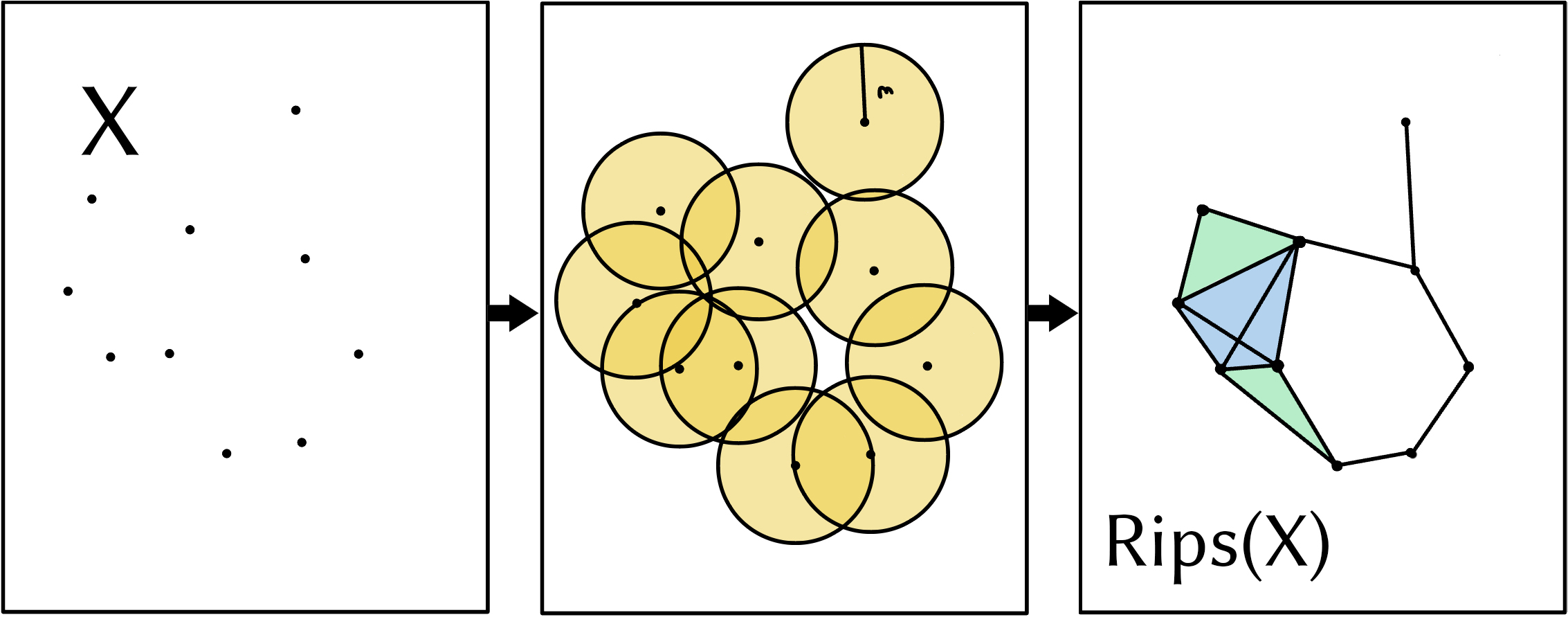}
    \caption{Process of Creating the Rips Complex Using a Set of Embeddings}
    \label{fig:Rips}
\end{figure}

\noindent $\epsilon>0$ is the radius of the open balls used to determine edge connections within the embedding space. For small values of $\epsilon$, the balls are too small to intersect and result in a complex with only the original embeddings. On the opposite end, if $\epsilon$ is too large, every feature in the input image is related to every other feature. We selected $\epsilon$ by performing a parameter study from $0.1$ to $1$ in increments of $0.05$. Our ideal value of $\epsilon$ captures the average number of features (i.e., UI components) on a given screen, which we found typically ranges from 20-100. \revision{The average number of triangles is determined using the Area-based Triangulated Embedding (ATE) method proposed by Krishna Vajjala \etal~\cite{vajjala2024vietoris}.} Through a small scale empirical study with two authors on a held out portion of our data, we found that this method of deriving triangle numbers struck a reasonable balance between sparsity and completeness that captured important feature-oriented UI component groups, minimizing less important connections between components. Through this process, we determined $\epsilon=0.5$ was most suitable, and produced an average of 100 2-simplices in the Rips complex. An illustration of the overlapping balls and the resulting simplices is shown in Fig.~\ref{fig:Rips}.%

By employing the Rips complex on the propagated UI embeddings \FRAME topologically captures the relationship between an object in an image with all other objects in the image as shown in Fig.~\ref{fig:Rips}. Once the simplex is created, we consider the embeddings that make $2$-simplices (2D triangles) and use the area of the resulting simplex as a weight for the involved embeddings. The area of a given 2-simplex can be calculated using Heron's formula~\cite{weisstein2003heron}:
\begin{equation}
    \text{Area} = \sqrt{S(S-A)(S-B)(S-C)}
\end{equation}
\noindent where $S$ is the semi-perimeter of the 2-simplex and $A,B,$ and $C$ are the lengths of its three edges. The area is then used to weigh the involved embeddings. \arun{Unlike simple pairwise distances (1-simplices) which often capture noisy, unrelated nearby elements, the Vietoris-Rips complex identifies cohesive neighborhoods that accurately represent distinct UI features. We use the 2-simplex (triangle) area for weighting because components consuming more screen real-estate are typically perceived as more ``functionally dominant'' and semantically important within the interface's layout.}

\FRAME then employs the Area-based Triangulated Embedding method (ATE), proposed by Krishna Vajjala \etal~\cite{vajjala2024vietoris}, for the weighted average of the embeddings. This creates a weighted relation between embeddings that are spatially close to each other, resulting in a weighted preservation of information between those three embeddings.
These embeddings are then concatenated together to create an embedding for an entire screen, as described in the next subsection. This process is shown in Fig.~\ref{fig:overview}-\circled{4}. 

\subsection{Embedding Post-Processing}

The process of creating a distance-similar graph, propagating the embeddings within it, and consolidating them using the area-based triangulation creates a semantically enhanced set of embeddings that encode the relationships of UI components which represent screen features. To create its final screen embedding, \FRAME combines the graph-based semantically enhanced embeddings with a CLIP embedding of the original greyscale and contrast enhanced input image, in order to capture general global image features, resulting in a 1,792-dimension embedding. \FRAME then further processes these embeddings using Principal Component Analysis (PCA)~\cite{PCA} to reduce potential noise in the resulting larger embedding. \FRAME uses Scikit-Learn's implementation of PCA applied to the training splits of our three experimental datasets, and the resulting \FRAME embeddings were reduced to a dimensionality of 116. To ensure that arbitrary future embeddings result in size 116 without re-computing the PCA, \FRAME uses PCA's transformation matrix and performs a dot product with the derived 1,792 dimensional vector for future images.

\section{Design of Empirical Evaluation}
\label{sec:eval}

\newlist{questions}{enumerate}{2}
\setlist[questions,1]{label=RQ\arabic*.,ref=RQ\arabic*}
\setlist[questions,2]{label=(\alph*),ref=\thequestionsi(\alph*)}
In this section, we describe the procedure we used to evaluate \FRAME. To achieve our study goals, we formulated the following
four research questions (RQs):

\begin{itemize}
	\item{\textbf{RQ$_1$} \textit{How does \FRAME perform against baselines in screen retrieval tasks?}}
    \item{\textbf{RQ$_2$} \textit{How does \FRAME perform against baselines in screen clustering tasks?}}
    \item{\textbf{RQ$_3$} \textit{How do \FRAME's embedding propagation, weighting, and image processing contribute to its effectiveness?}}
    \item{\textbf{RQ$_4$} \textit{Why does \FRAME outperform baselines in screen retrieval?}}
    
\end{itemize}

\subsection{Evaluation Datasets}

We focus \FRAME's evaluation on screen \textit{retrieval} and \textit{clustering} as these tasks have been shown to be important in software engineering tools/tasks~\cite{Behrang_18,Cardenas:ICSE'19,saha2024screenbug,Cooper:ICSE'21,Yan:ICSE'24}.

We evaluate \FRAME with three different datasets derived from prior work that use semantic screen similarity to aid in UI testing (\eg the Avgust~\cite{Avgust} and Aurora~\cite{khan2024aurora} datasets) and UI design (\eg the Enrico~\cite{Leiva20_enrico} dataset).

\subsubsection{Aurora Dataset} 
The Aurora dataset, created by Khan \etal~\cite{khan2024aurora}, was derived to support a UI testing tool that classified UI screens into various design patterns so that it can intelligently test the identified screens. This dataset was derived by sampling 1\% of the images from the larger RICO~\cite{Rico} dataset, and clustering them to bootstrap a labeling process. The authors of the paper then manually grouped 1,370 UI screens into 21 semantic categories, using a rigorous, independent labeling procedure with high agreement~\cite{khan2024aurora}.
For our retrieval task, we perform a stratified sampling of 10\% across categories to serve as screen queries, whereas the other 90\% of the dataset represents the corpus. 

\subsubsection{Avgust Dataset}
The Avgust dataset was introduced by Zhao \etal~\cite{Avgust}, also in support of automated cross-app mobile testing. It has 25 labels corresponding to the purpose of a given screen (\eg shopping cart) across 2,475 total screens. This dataset was manually labeled by four authors who mutually agreed on each screen's category. This is a larger dataset with approximately 99 screens per category on average. We again use the same 90/10 split for corpus/queries.

\subsubsection{Enrico Dataset}
The Enrico dataset is designed for topic classification tasks for UI screens to assist in UI Design. It consists of 1,460 unique UI screens labeled by their design categories (\eg Profile screen) by two authors. The screens span 20 labels, creating an uneven distribution of class support with the largest class having 265 screens and the smallest class having only nine screens. We use an 85/15 corpus/query split for this dataset given its smaller size. The Enrico dataset is perhaps the most challenging to classify or cluster due to its small support across screen categories.

\revision{These datasets were selected for their diversity and relevance to UI programming tools, such as automated UI test generation and UI design. Though originally developed for distinct end-to-end software engineering tasks, these datasets capture diverse functionalities and nuanced categorizations of UI screens. The carefully curated labels, established by multiple authors in their respective studies, offer a reliable benchmark to assess \FRAME's effectiveness in identifying specific screens for clustering and retrieval tasks. 
Notably, none of the images in Avgust, Aurora, and Enrico overlap, although some label categories recur due to the limited, finite existence of various app screen categories. This selection represents an extensive and diverse set of Android screens, supporting \FRAME's generalizability.}

\subsection{Baseline Embedding Techniques}

We compare \FRAME to four key baselines: CLIP, BLIP, Screen2Vec, and BERT. We use these baselines to measure \FRAME's efficacy compared to popular tools. We were not able to use more specialized baseline UI models such as ScreenAI~\cite{baechler2024screenai}, VUT~\cite{li2021vut}, UIBert~\cite{bai2021uibert}, or Graph4GUI~\cite{Jiang_2024}, as these techniques are not publicly available. However, it should be noted that \FRAME is largely complementary to these techniques. \noindent\revision{Additionally, this study aimed to provide a reproducible, open-source evaluation, which influenced the selection of the included state-of-the-art embedding techniques.}

\subsubsection{CLIP}

CLIP~\cite{clip} embeddings represent images and corresponding image keywords in a shared embedding space, with a contrastive learning mechanism that aims to minimize the distance between corresponding image-description pairs, enabling cross-modal understanding.

\subsubsection{BLIP}

Salesforce's BLIP embeddings~\cite{blip} extend the concept behind the multimodal shared embedding space of CLIP by using slightly different encoders and a feature that generates and aligns image captions for pre-training.

\subsubsection{Screen2Vec}

Li \etal's Screen2Vec~\cite{Li21} is designed as an embedding specifically for screens that captures and embeds information using runtime UI metadata and text.

\subsubsection{BERT}

Despite being primarily designed for text processing, BERT~\cite{BERT} can be re-purposed to encode the text displayed on UI screens.

We selected CLIP, BLIP, Screen2Vec, and BERT as our baseline models because each offers a different approach to embedding creation relevant to UI understanding: CLIP and BLIP provide multi-modal vision-language embeddings, Screen2Vec focuses on UI-specific text and hierarchy, and BERT captures textual information. We excluded larger LLMs such as GPT-4/5 from our retrieval/clustering tasks for three reasons: (i) it is difficult to represent screen-retrieval tasks via prompting techniques due to limitations with the context windows of LLMs, and most commercial frontier models do not offer APIs for generating image embeddings; (ii) image or text-based LLM-based embeddings tend to perform similarly or worse to vision-language models such as CLIP in general UI understanding tasks~\cite{Wu2024}, with LLMs performing better at processing UI Code~\cite{Duan2024}; (iii) computing large numbers of embeddings for practical applications of semantic screen search using LLMs is likely to be costly.

\subsection{\textbf{RQ$_1$} - Screen Retrieval Methodology}
\label{sec:eval_metrics}
To evaluate \FRAME for the task of screen retrieval, we formulate a query with each UI screen in a held out query set for each dataset, and calculate the cosine similarity, using a given embedding technique, against each screen in the corpus set, forming a ranked list of retrieved screens. We then adopt the widely used MRR and HR@\textit{k} (Hit Rate) metrics~\cite{Mahmud:ICSE'24,saha2024screenbug} to evaluate the effectiveness of \FRAME compared to our baseline techniques. HR@\textit{k} measures the percentage of queries for which an approach retrieves at least one correct screen within the top-\textit{k} screens. Here a \textit{correct screen} is one that shares the same label as the query. MRR calculates the average of the reciprocals of ranks at which the relevant items are retrieved. Higher numbers indicate better performance across metrics.

\begin{table}[t]
\footnotesize
\centering
\vspace{-1em}
\caption{\FRAME Pre-Processing Ablation Study Configurations}
\vspace{-1em}
\begin{tabular}{|c|c|c|}
\hline
\textbf{Acronym}   & \textbf{Greyscale} & \textbf{Contrast} \\ \hline
\textbf{NG-C}  &               \xmark             &           \cmark                         \\ \hline
\textbf{NG-NC}  &                \xmark            &            \xmark                      \\ \hline
\textbf{G-NC}  &                \cmark            &            \xmark                        \\ \hline

\end{tabular}
\label{tab:PreProAblation}
\end{table}

\subsection{\textbf{RQ$_2$} - Screen Clustering Methodology}
To answer this RQ we generate \FRAME and CLIP embeddings across the three datasets and evaluate their clustering performance (using K-means clustering, with k set to the number of labels within each dataset) across nine common clustering metrics: Sum of Squared Errors (SSE), Clustering Accuracy, Fowlkes-Mallows Index (FMI), Homogeneity, Completeness, V-Measure, Davies-Bouldin Index (DBI), Adjusted Rand Index (ARI), and Normalized Mutual Information (NMI). This experiment allows us to examine how well \FRAME is able to segment groups of screens using its enhanced embeddings compared to CLIP, which was the best performing baseline in the screen retrieval task.

\subsection{\textbf{RQ$_3$} - Ablation Study Methodology}  
The goal of this experiment is to determine whether \FRAME's various features and pre-processing techniques contribute to its effectiveness in screen retrieval. We divide the ablation study into two parts: evaluating pre-processing configurations and analyzing \FRAME's specific embedding components. We utilize only the Avgust and Aurora datasets for this evaluation, excluding Enrico due to its inherent class imbalance and small sample size.

The first part of the study examines \FRAME's greyscale and contrast image pre-processing in conjunction with all model components. As shown in Table~\ref{tab:PreProAblation}, we evaluate three configurations: (i) \texttt{\small\textbf{NG-C}}, representing contrast without greyscale; (ii) \texttt{\small\textbf{NG-NC}}, which includes neither augmentation; and (iii) \texttt{\small\textbf{G-NC}}, representing greyscale without contrast.

The second phase evaluates embedding variants using a three-tuple notation, as detailed in Table \ref{tab:Ablation}. For example, in the tuple \texttt{\small\textbf{C-PB-PC}}, the first grouping indicates the use of the full-screen CLIP embedding (\texttt{\small\textbf{C}} or \texttt{\small\textbf{NC}}); the second grouping denotes the use of propagated BERT-based component text embeddings (\texttt{\small\textbf{PB}} or \texttt{\small\textbf{NPB}}); and the third grouping determines whether propagated and weighted CLIP embeddings are utilized (\texttt{\small\textbf{PC}} or \texttt{\small\textbf{NPC}}) with \texttt{\small\textbf{N}} indicating \textit{not present}.

\begin{table}[t]
\footnotesize
\vspace{-1em}
\caption{\FRAME Ablation Study Configurations}
\vspace{-1em}
\begin{tabular}{|c|c|c|c|}
\hline
\textbf{Acronym}   & \textbf{Whole Screen CLIP} & \textbf{Node BERT} & \textbf{Node CLIP} \\ \hline
\textbf{NC-PB-PC}  &               \xmark             &           \cmark         &  \cmark                  \\ \hline
\textbf{C-PB-NPC}  &                \cmark            &            \cmark        &   \xmark                 \\ \hline
\textbf{C-NPB-PC}  &                \cmark            &            \xmark        &   \cmark                 \\ \hline
\textbf{C-NPB-NPC} &                    \cmark        &             \xmark       &   \xmark                 \\ \hline
\textbf{NC-PB-NPC} &                    \xmark        &             \cmark       &    \xmark                \\ \hline
\textbf{NC-NPB-PC} &                 \xmark           &             \xmark       &    \cmark                \\ \hline
\end{tabular}
\label{tab:Ablation}
\end{table}

\vspace{-3pt}
\subsection{\textbf{RQ$_4$} - Qualitative Study on Screen Retrieval}
To evaluate the performance of \FRAME in comparison to CLIP embeddings, we conduct a qualitative analysis using the Avgust dataset. We randomly select a test image and retrieve the three most similar images based on the embeddings from each model. By visually inspecting these top matches, we assess the ability of each embedding to capture and represent the visual and structural characteristics of UI screens accurately.

\section{Empirical Results}
\label{sec:results}

\subsection{\textbf{RQ$_1$}: \textit{\FRAME Screen Retrieval Effectiveness}}

\noindent \FRAME performs significantly better in screen retrieval tasks compared to the state-of-the-art baseline methods, as shown in Table~\ref{MainResults}. Specifically, \FRAME outperforms Screen2Vec, BERT, BLIP, and CLIP in every metric across all three datasets.

\FRAME significantly outperforms baselines across three diverse datasets, exhibiting an average improvement of 16.0\% on the sparse Aurora dataset and 4.9\% on the Avgust dataset. Despite the increased difficulty of the Enrico dataset characterized by class imbalance, \FRAME maintains superior performance with minor increases in hit-rate and MRR. Specifically, compared to the strongest baseline, CLIP, \FRAME achieved a 12.8\% MRR improvement on Aurora and a 3.0\% improvement on Avgust, with most results across all three benchmarks reaching statistical significance. Collectively, these results over highly varied datasets demonstrate \FRAME's effectiveness in accurately ranking similar UI screens toward the top of retrieval lists.

\begin{table}[t]
\centering
\vspace{-1em}
\caption{Retrieval results over 3 datasets, where the best results are in \textbf{bold}, and \textbf{*} indicates a $p$-value less than 0.05 from a two-tailed paired t-test between \FRAME and the best baseline.}
\renewcommand{\arraystretch}{1}
\scalebox{0.62}{
\begin{tabular}{|l|c|c|c|c|c|c|c|}
\toprule
\textbf{Dataset} & \textbf{Metrics} & \textbf{Screen2Vec} & \textbf{BERT} & \textbf{BLIP} & \textbf{CLIP} & \textbf{FRAME} & \textbf{$p$-value} \\
\midrule
\multirow{4}{*}{Aurora} & HR@1 & 0.3750 & 0.3194 & 0.4097 & 0.4722 & \textbf{0.5625*} & $8.56 \times 10^{-15}$ \\
& HR@5 & 0.3027 & 0.2486 & 0.3944 & 0.4013 & \textbf{0.4652*} & $6.02 \times 10^{-22}$ \\
& HR@10 & 0.2604 & 0.2187 & 0.3639 & 0.3590 & \textbf{0.4166*} & $3.38 \times 10^{-26}$ \\
& MRR & 0.4853 & 0.4363 & 0.5479 & 0.5963 & \textbf{0.6729*} & $5.09 \times 10^{-14}$ \\
\hline
\multirow{4}{*}{Avgust} & HR@1 & 0.3922 & 0.8270 & 0.8509 & 0.8549 & \textbf{0.8823*} & $2.90 \times 10^{-10}$ \\
& HR@5 & 0.3185 & 0.7788 & 0.7937 & 0.7788 & \textbf{0.8329*} & $7.41 \times 10^{-23}$ \\
& HR@10 & 0.2739 & 0.7141 & 0.7522 & 0.7364 & \textbf{0.7827*} & 0.00150 \\
& MRR & 0.5116 & 0.8908 & 0.8861 & 0.8913 & \textbf{0.9184*} & $9.42 \times 10^{-10}$ \\
\hline
\multirow{4}{*}{Enrico} & HR@1 & 0.2895 & 0.2212 & 0.2837 & 0.3365 & \textbf{0.3894*} & 0.0176 \\
& HR@5 & 0.2389 & 0.2019 & 0.2712 & 0.3048 & \textbf{0.3096} & 0.0954 \\
& HR@10 & 0.2063 & 0.1822 & 0.2577 & 0.2740 & \textbf{0.2903} & 0.0687 \\
& MRR & 0.4271 & 0.3849 & 0.4534 & 0.5051 & \textbf{0.5391} & 0.0586 \\
\bottomrule
\end{tabular}
}
\label{MainResults}
\end{table}

In order to better understand the advantages in performance that \FRAME offers, we manually examined both the metrics and the lists of retrieved screens for the HR@10 metric between \FRAME and CLIP, the best performing baseline, across individual screen categories across the datasets. We found that \FRAME consistently outperformed CLIP across all screen types except for four of Avgust's twenty-five different screen types: \textit{SignIn}, \textit{Menu}, \textit{SignUp}, and \textit{Account} screens. We speculate that \FRAME's lower accuracy on these screens stems from their inherently high variability, reliance on textual semantics, and flexible layouts, which favor CLIP's multi-modal understanding. In particular, \textit{SignIn} and \textit{SignUp} screens often share similar structures and input fields, making it challenging for \FRAME to distinguish them based on functionality alone.

\subsection{\textbf{RQ$_2$}: \textit{\FRAME Screen Clustering Effectiveness}}
Our results suggest that \FRAME embeddings outperform CLIP embeddings across almost all metrics, except for SSE, in the Enrico and Avgust datasets, as seen in Table~\ref{tab:clusterres}. Specifically, \FRAME achieves a Clustering Accuracy of 0.0543 on the Enrico dataset and 0.1669 on the Avgust dataset, compared to CLIP's 0.0074 and 0.0072, respectively. This shows that \FRAME performs better at correctly grouping similar screens.

\begin{table}[t]
\small
\renewcommand{\arraystretch}{1}
\setlength{\tabcolsep}{4pt}
\centering
\vspace{-1em}
\caption{Comparison of \FRAME and CLIP Using Clustering Metrics Across Three Datasets}
\vspace{-0.5em}
\begin{tabular}{l|c|c|c}
\multicolumn{4}{c}{\textbf{Enrico Data}} \\
\hline
Embedding & SSE & Clustering Accuracy & FMI \\
\hline
CLIP & \textbf{9832.0852} & 0.0074 & 0.1372 \\
FRAME & 68604.9113 & \textbf{0.0543} & \textbf{0.2314} \\
\hline
  & Homogeneity & Completeness & V-Measure \\
\hline
CLIP & 0.0083 & 0.0081 & 0.0082 \\
FRAME & \textbf{0.1364} & \textbf{0.1362} & \textbf{0.1363} \\
\hline
  & DBI & ARI & NMI \\
\hline
CLIP & 6.8759 & -0.0025 & 0.0082 \\
FRAME & \textbf{2.7673} & \textbf{0.1012} & \textbf{0.1363} \\
\hline
\multicolumn{4}{c}{\textbf{Avgust Data}} \\
\hline
Embedding & SSE & Clustering Accuracy & FMI \\
\hline
CLIP & \textbf{17044.7382} & 0.0072 & 0.1394 \\
FRAME & 54236.0280 & \textbf{0.1669} & \textbf{0.4170} \\
\hline
  & Homogeneity & Completeness & V-Measure \\
\hline
CLIP & 0.0112 & 0.0107 & 0.0109 \\
FRAME & \textbf{0.4742} & \textbf{0.4668} & \textbf{0.4705} \\
\hline
  & DBI & ARI & NMI \\
\hline
CLIP & 7.2990 & 0.0032 & 0.0109 \\
FRAME & \textbf{2.1801} & \textbf{0.3183} & \textbf{0.4705} \\
\hline
\multicolumn{4}{c}{\textbf{Aurora Data}} \\
\hline
Embedding & SSE & Clustering Accuracy & FMI \\
\hline
CLIP & 86165.4671 & 0.0017 & 0.1403 \\
FRAME & \textbf{47169.3284} & \textbf{0.0217} & \textbf{0.2799} \\
\hline
  & Homogeneity & Completeness & V-Measure \\
\hline
CLIP & 0.0192 & 0.0208 & 0.0199 \\
FRAME & \textbf{0.2530} & \textbf{0.2702} & \textbf{0.2613} \\
\hline
  & DBI & ARI & NMI \\
\hline
CLIP & 11.5100 & -0.0015 & 0.0199 \\
FRAME & \textbf{3.5316} & \textbf{0.1637} & \textbf{0.2613} \\
\hline
\end{tabular}
\label{tab:clusterres}
\end{table}

\begin{figure}[t]
    \centering
    \vspace{-1em}
    \begin{minipage}[b]{1.0\linewidth}
        \centering
        \includegraphics[width=\linewidth]{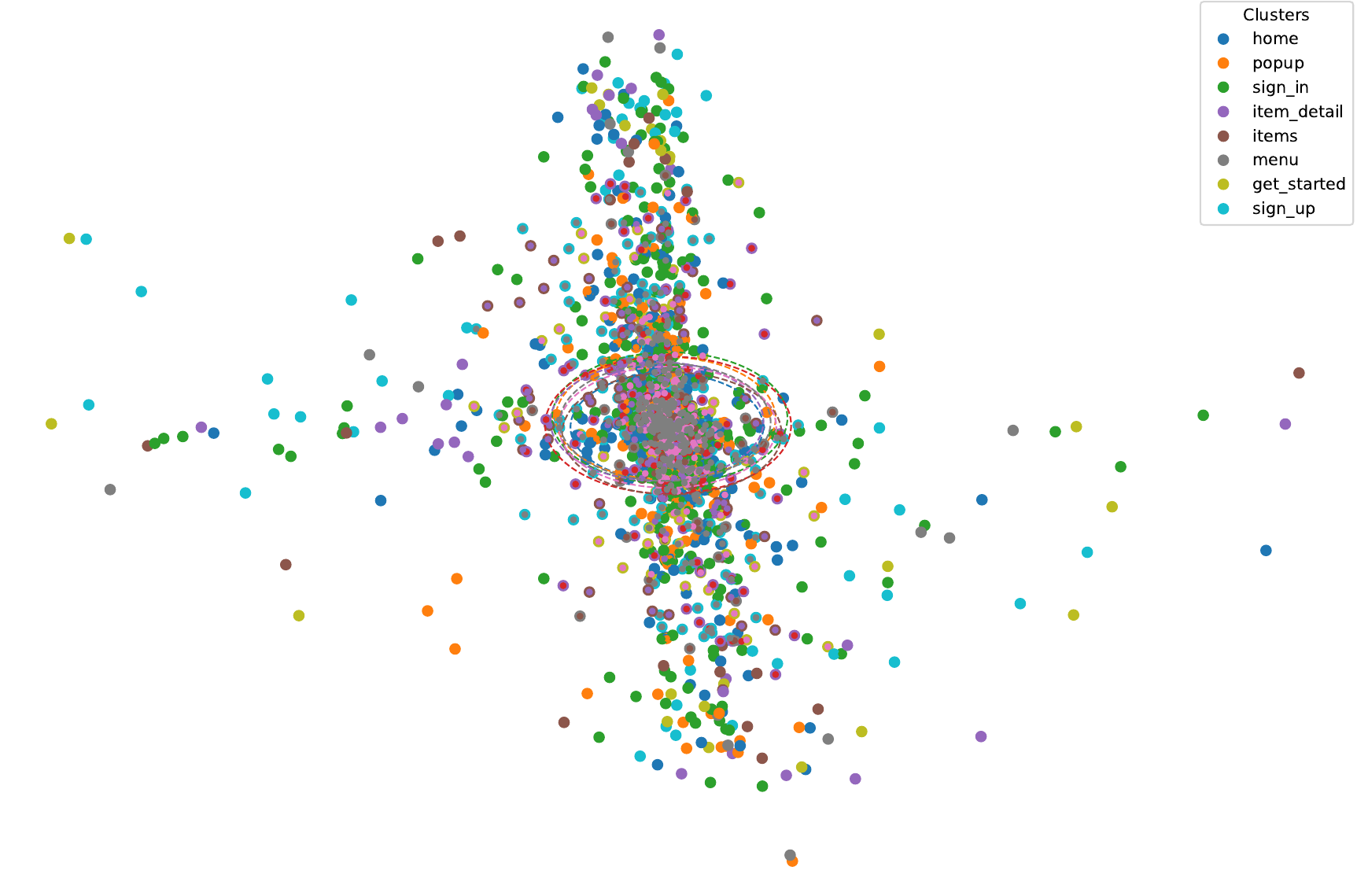}
        \vspace{-1.5em}
        \caption{CLIP Embedding Clustering -- Avgust Dataset}
        \label{fig:clipcluster}
    \end{minipage}
    \vspace{0.2cm}
    \begin{minipage}[b]{1.0\linewidth}
        \centering
        \includegraphics[width=\linewidth]{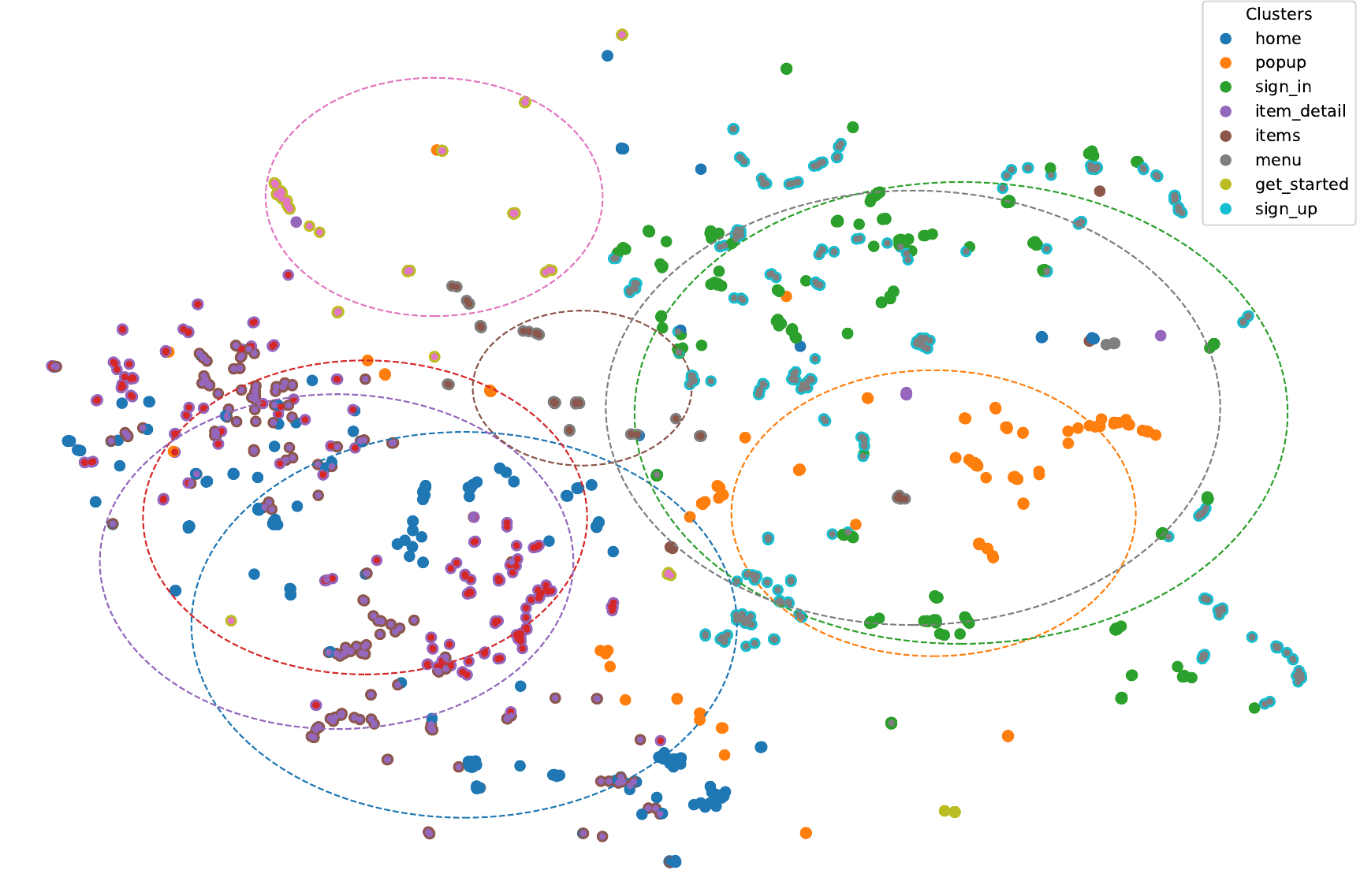}
        \vspace{-1em}
        \caption{\FRAME Embedding Clustering -- Avgust Dataset}
        \vspace{-1.5em}
        \label{fig:framecluster}
    \end{minipage}
    \label{fig:vertical_images}
\end{figure}

\FRAME outperforms CLIP across both internal and external clustering metrics, achieving significantly higher FMI (0.4170 vs. 0.1394), V-Measure (0.4705 vs. 0.0109), and ARI (0.3183 vs. 0.0032) on the Avgust dataset. These results, along with high homogeneity and completeness scores, present \FRAME's ability to produce clusters aligned with true functional structures. While CLIP achieves lower SSE—indicating tighter clusters—this tightness sacrifices accuracy through over-generalization. In contrast, \FRAME's lower DBI values (e.g., 2.1801 vs. 7.2990 for Avgust) confirm better cluster separation and a refined ability to distinguish unique UI~features.

Visualizations in Figs.~\ref{fig:clipcluster} and~\ref{fig:framecluster} show that CLIP creates tighter, more homogeneous clusters that sacrifice accuracy due to a lack of structural granularity. In contrast, \FRAME effectively identifies and groups functionally similar screens despite minor outliers. While 2D projections limit feature detail, the quantitative data in Table~\ref{tab:clusterres} supports these visual insights, confirming \FRAME's superior ability to differentiate UI features.

\subsection{\textbf{RQ$_3$}: \textit{\FRAME Component Ablation Study}}

\begin{table*}[h]
\centering
\small
\vspace{-1em}
\caption{Component ablation over 2 datasets between \FRAME and its variants, where the best results are in \textbf{bold}.}
\vspace{-1em}
\renewcommand{\arraystretch}{1}

\begin{tabular}{|c|c|c|c|c|c|c|c|c|c|c|}
\toprule
\textbf{Dataset} & \textbf{Metrics} & \textbf{NC-NPB-PC} & \textbf{NC-PB-NPC} & \textbf{C-NPB-NPC} & \textbf{NC-PB-PC} & \textbf{C-PB-NPC} & \textbf{C-NPB-PC} & \textbf{FRAME}\\
\hline
\midrule
\multirow{4}{*}{Aurora} 
& HR@1 & 0.2777 & 0.3750 & 0.4722 & 0.4166 & 0.5138 & 0.4513 & \textbf{0.5625}\\
& HR@5 & 0.1944 & 0.2972 & 0.4013 & 0.2750 & 0.4486 & 0.4180 & \textbf{0.4652}\\
& HR@10 & 0.1631 & 0.2791 & 0.3590 & 0.2458 & \textbf{0.4180} & 0.3618 & 0.4166\\
& MRR & 0.3880 & 0.5026 & 0.5963 & 0.5136 & 0.6448 & 0.6077 &\textbf{0.6729} \\
\hline
\hline
\multirow{4}{*}{Avgust} 
& HR@1 & 0.7960 &  0.8352 & 0.8549 & 0.8666 & 0.8588 & 0.8705 & \textbf{0.8823}\\
& HR@5 & 0.6901 &  0.7850 & 0.7788 & 0.8023 & 0.8282 &  0.8219 & \textbf{0.8329}\\
& HR@10 & 0.5929 & 0.7360 &  0.7364 & 0.7498 & 0.7721 & 0.7674 & \textbf{0.7827} \\
& MRR & 0.8380 &  0.8724 & 0.8913 & 0.9012 & 0.9014 & 0.9114 & \textbf{0.9184}\\
\bottomrule
\end{tabular}
\label{AblatioTable}
\end{table*}
\begin{table}[t!]
\centering
\footnotesize
\vspace{-1.5em}
\caption{Pre-processing ablation over 2 datasets between \FRAME and its greyscale/contrast variants, where the best results are in \textbf{bold}.}
\vspace{-1em}
\renewcommand{\arraystretch}{1}
\begin{tabular}{|c|c|c|c|c|c|}
\toprule
\textbf{Dataset} & \textbf{Metrics} & \textbf{NG-C} & \textbf{NG-NC} & \textbf{G-NC} & \textbf{FRAME} \\
\hline
\midrule
\multirow{4}{*}{Aurora} 
& HR@1 & 0.4930 & 0.5347 & 0.5208 & \textbf{0.5625} \\
& HR@5 & 0.4069 & 0.4250 & 0.4527 & \textbf{0.4652} \\
& HR@10 & 0.3638 & 0.3805 & 0.4138 & \textbf{0.4166} \\
& MRR & 0.6171 & 0.6484 & 0.6492 & \textbf{0.6729} \\
\hline
\hline
\multirow{4}{*}{Avgust} 
& HR@1 & 0.8505 & 0.8784 & 0.8784 & \textbf{0.8823} \\
& HR@5 & 0.8119 & 0.8298 & 0.8266 & \textbf{0.8329} \\
& HR@10 & 0.7658 & \textbf{0.7827} & 0.7799 & \textbf{0.7827} \\
& MRR & 0.9094 & 0.9118 &  0.9138 &  \textbf{0.9184} \\
\bottomrule
\end{tabular}
\label{AugmentationTable}
\end{table}

\subsubsection{Pre-Processing Ablation Study}
We first present the results for the pre-processing configurations in Table~\ref{AugmentationTable}. The \textbf{NG-C} (contrast only) variant is the least effective configuration, ranking last on all eight dataset--metric combinations. It is also outperformed by \textbf{NG-NC} (no augmentation) on every metric, indicating that raising contrast on a full-color screenshot actively hinders component detection rather than aiding it. The \textbf{G-NC} (greyscale only) variant is the strongest single-augmentation configuration, and the best alternative to the full model on Aurora, where it leads \textbf{NG-NC} on three of four metrics. On Avgust the two are comparable: \textbf{G-NC} ties on HR@1 and leads on MRR, but trails slightly on HR@5 and HR@10. This suggests that removing color bias is most beneficial on visually diverse corpora such as Aurora. Overall, greyscale conversion acts as a prerequisite for contrast enhancement to be useful -- contrast alone is detrimental, whereas combining the two yields the best or tied-best result on all eight measurements.

\subsubsection{Embedding Ablation Study}
Table~\ref{AblatioTable} details the performance of \FRAME's various component configurations. The full \FRAME variant, utilizing all embeddings, outperforms all others except for HR@10 on the Aurora dataset. Variants lacking the global screen-level CLIP embedding (\texttt{\small\textbf{NC-NPB-PC}}, \texttt{\small\textbf{NC-PB-NPC}}, and \texttt{\small\textbf{NC-PB-PC}}) show poor performance on the Aurora dataset, as they rely solely on localized component propagation without broader screen context. While incorporating screen-level context (\texttt{\small\textbf{C-NPB-PC}} and \texttt{\small\textbf{C-PB-NPC}}) improves results, these configurations still underperform relative to the full model. These results confirm that embedding propagation is essential for capturing cohesive UI screen semantics.

\subsection{\textbf{RQ$_4$}: \textit{Qualitative Analysis}}
\begin{figure}[t]
    \centering
    \begin{minipage}[b]{1.0\linewidth}
        \centering
        \includegraphics[width=\linewidth]{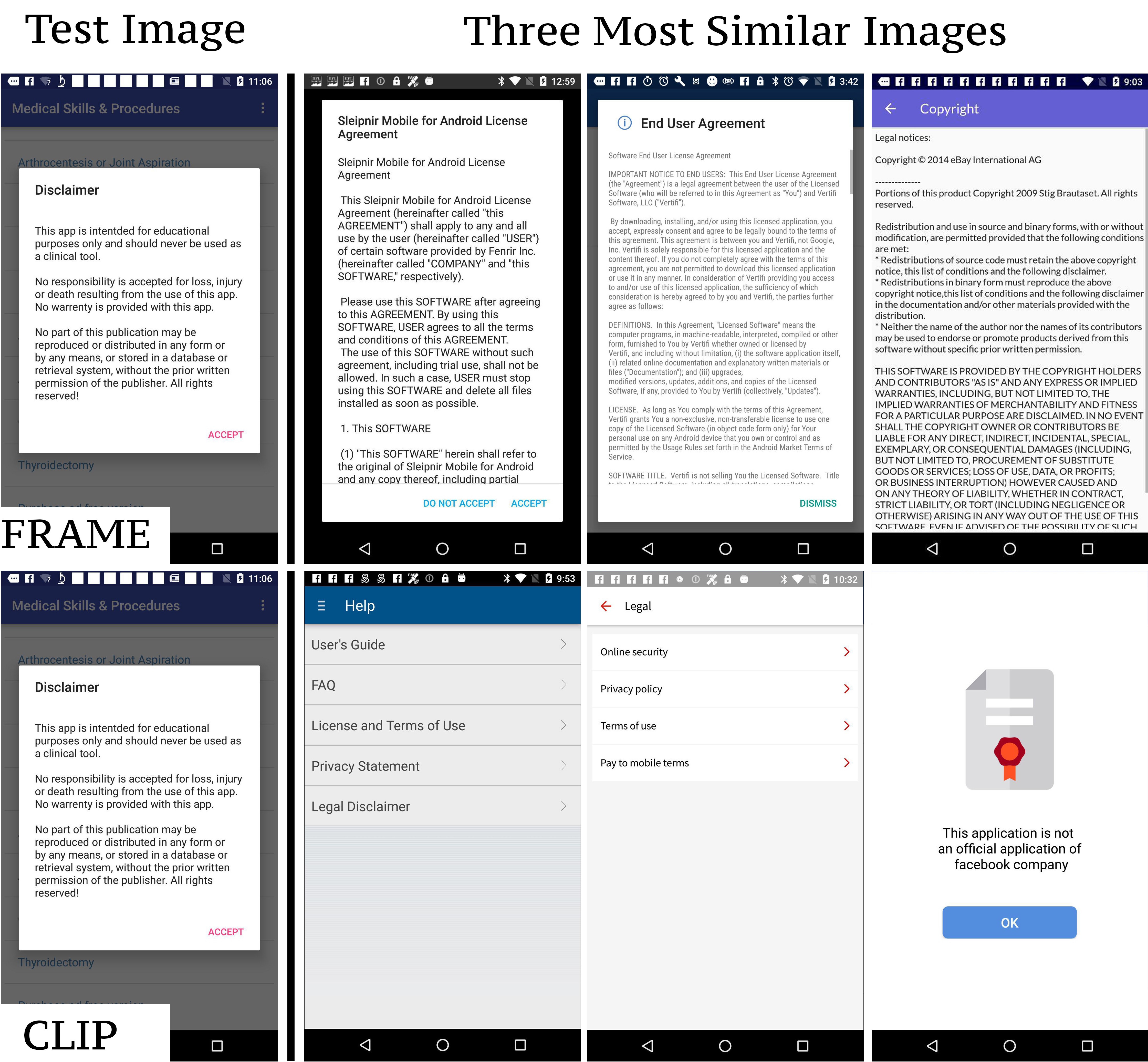}
    \end{minipage}
    \vspace{-1em}
    \caption{Comparison of the Top-3 Most Similar Screens for Each Embedding Type}
    \label{fig:modals}
\end{figure}

In this section we illustrate the qualitative benefits that \FRAME provides in regards to screen structure. Fig.~\ref{fig:modals} offers insight into how various models identify similar screens. The screen on the left is the test image and the three images on the right are the three most similar screens to the test image using the \FRAME and CLIP embeddings. Similarity is computed using cosine similarity. 

The evaluation of modal screens shown in Fig.~\ref{fig:modals} using the embeddings further reinforces the strengths of \FRAME in identifying similar images based on component relationships. \FRAME's selected images for modal screens show strong functional similarity, including consistent modal placement and content within the modal. This highlights \FRAME's capability to recognize and prioritize the overall feature set of modal screens, making it suitable for tasks requiring transferable understanding of UI screens across UI design variations. In comparison, CLIP's retrieved screens are far less related to the query, and tend to capture general patterns, instead of features indicated by UI component relationships.

\section{Discussion}
\label{sec:discussion}

\subsection{Analysis of Retrieval Failures}
\arun{While \FRAME consistently outperformed CLIP, we observed lower accuracy on four specific screen types: \textit{SignIn}, \textit{Menu}, \textit{SignUp}, and \textit{Account}. We attribute this to the inherent structural similarity between \textit{SignIn} and \textit{SignUp} screens, which often share near-identical input field layouts. In these cases, the model relies heavily on textual semantics, where CLIP's multimodal pre-training offers a slight advantage over \FRAME's structural priors.}

\subsection{The ``Tightness vs. Accuracy'' Trade-off}
\arun{A key finding in our clustering evaluation is the discrepancy between SSE and other performance metrics. While CLIP generally achieves lower SSE (indicating ``tighter'' clusters), this tightness is often misleading. Visualizations in Figs.~\ref{fig:clipcluster} and~\ref{fig:framecluster} reveal that CLIP-based embeddings force diverse screens into a small, overlapping area of the embedding space. This over-generalization sacrifices the granularity needed to distinguish functional features. In other words, while clusters are tighter, they occupy the same semantic embedding space, making it difficult to discern discrete types of screens. In some cases, this could also lead to a deceptively high accuracy, where a majority of screens fit within the area of overlap of these small clusters. However, this is not a good representation of the discriminatory power of the baseline model, as indicated by our other measured metrics. In contrast, \FRAME's lower DBI values (2.1801 vs.\ CLIP's 7.2990) indicate better-separated clusters that align more closely with true structural features.}

\subsection{Neuro-Symbolic vs. Language/Vision Model-Based Representations}
\arun{The comparison with BERT/CLIP suggests that pure LM-based or VLM-based textual descriptions lack the structural precision captured by \FRAME's approach. While LMs can be adept at high-level categorization, they often omit the spatial relationships and component neighborhoods that define UI features. \FRAME’s use of graph propagation and geometric weighting allows it to maintain structural fidelity that generalized descriptions cannot match.}

\section{Related Work \& Novelty Statement}
\label{sec:rel-work}

In this section, we discuss existing multimodal UI learning techniques, and discuss their limitations in context, differentiating and explicitly illustrating \FRAME's novelty.

\subsection{UI Learning Models}
Baechler \etal~\cite{baechler2024screenai} proposed ScreenAI, a Vision-Language Model for UI comprehension. ScreenAI integrates vision and language processing to understand and interact with UIs and infographics. ScreenAI makes use of a text-based representation of screens and a novel UI component labeling task to excel in screen question answering. However, this model does not directly encode UI relationships and is not immediately applicable to the retrieval and clustering tasks \FRAME targets. Li \etal~\cite{li2021vut} proposed VUT, a UI transformer for multimodal, multi-task user interface modeling. This approach considers the hierarchical structure, image, and language properties of the screen to create their screen representations. Bai \etal~\cite{bai2021uibert} proposed UIBert, a transformer-based joint image-text model that learns generic feature representations for a UI and its components. This embedding is similar to VUT in its goal to create an embedding using image, text, and structural screen properties. Screen2Vec~\cite{Li21}, UIBert~\cite{bai2021uibert}, and VUT~\cite{li2021vut} do not directly capture the spatial and functional relationships between UI components, often reducing structural information to flat representations. This limits their effectiveness in tasks such as duplicate detection, clustering, and app categorization, where understanding component interdependence is crucial. 

The closest related work to our own was introduced by Jiang \etal~\cite{Jiang_2024} in Graph4GUI, a graph neural network-based representation of UIs. This graph comprises element nodes, representing properties like appearance and size, and constraint nodes, representing layout constraints such as alignment and grouping. However, this model is tailored for UI generation, and represents the \textit{structural} properties of UI components, as opposed to attempting to capture \textit{feature-oriented} component relationships.
It is important to note that \textit{\textbf{the only publicly available model discussed is Screen2Vec~\cite{Li21}}}, and further each model tends to target different UI tasks, making them difficult to apply to screen$\leftrightarrow$screen similarity calculations.

\subsection{UI Search and Similarity}
Guigle, proposed by Bernal-C\'ardenas \etal~\cite{Cardenas:ICSE'19}, is a UI search engine. It uses metadata information to facilitate a filter-based search, rather than an embedding-based search. Behrang \etal~introduced GuiFetch~\cite{Behrang_18}, which computes similarity scores at the app level with UI sketches using keywords and GUI hierarchy properties. GeminiScope, proposed by Mao \etal~\cite{Mao18}, is a GUI similarity metric that uses the leaf nodes in the hierarchy tree to determine the position of the elements on the screen. They compute the absolute values of UI-related features, such as position and size, and use these metrics as their numerical representation for the screen.
However, GeminiScope does not consider relations between the embeddings.

\subsection{UI Comprehension}
To modify or test a UI, it is necessary to have a semantic understanding of the UI. Many tools use the Android screen hierarchy~\cite{bai2021uibert,li2021vut,Li21,Schoop22}, however, the hierarchy is not always available. Wu \etal~introduce an approach called Screen Parsing~\cite{Wu:UIST'21} to reverse engineer the screen hierarchy using computer vision techniques. Spotlight, introduced by Li \etal~\cite{li20spotlight}, uses computer vision to create a vision-language model architecture that can assist in screen comprehension tasks. \FRAME differs from both of these techniques in both its goal and its UI representation.
\vspace{0.5em}%

\noindent\textbf{Novelty Statement:} In summary, \FRAME differs from the techniques discussed in the prior three subsections in a number of key ways. First, it is the first \textit{feature-oriented} approach for calculating UI similarity, combining both generalized component representations using VLMs, and a symbolic representation of UI structure in a graph form for encoding features. Second, \FRAME is the first graph-based technique to leverage embedding propagation and weighting techniques that capture explicit UI component relationships representing features. While some other techniques (\eg Graph4GUI) encode hierarchy information into a graph, this information is largely structural and not captured to represent \textit{features}. Third, \FRAME is an \textit{enhancement} for existing large vision-language models that does not require additional training, while taking advantage of the continuing rapid advancements of such models. Fourth, our current implementation of \FRAME does not require additional model training, making use of strong UI priors to effectively capture UI information. Together, these properties constitute \FRAME's novelty.

\vspace{-0.5em}\section{Conclusions}\vspace{-0.5em}
\label{sec:conclusion}

In this paper, we presented \FRAME, a neuro-symbolic approach for creating UI screen embeddings that excel in computing semantic screen similarity. We measured the effectiveness and generalizability of \FRAME on various open source UI datasets in comparison to strong baselines. Our results indicate that \FRAME is effective and outperforms key baselines, offering a novel approach for creating embeddings for UI similarity. \FRAME is constructed by design as an extensible model that can take advantage of ever-improving large language and vision model representations.

\section*{Data Availability}
The code and data associated with this paper along with the experimental infrastructure for evaluating \FRAME are provided in an archived replication package~\cite{REPO} and an accompanying source code repository~\cite{REPO-GH} to facilitate replication and future work on the topics of screen retrieval and clustering.

\section*{Acknowledgments}
This research is supported in part by NSF grants CCF-2441355 and CCF-2311469. Any opinions, findings, and conclusions expressed herein are the authors' and do
not necessarily reflect those of the sponsors.

\balance

\bibliographystyle{IEEEtran}
\bibliography{bibliography,ref}

\end{document}